\documentclass[format=sigconf, review=false, anonymous=false, screen, dvipsnames]{lib-acm/acmart}

\setcopyright{acmlicensed}
\copyrightyear{2026}
\acmYear{2026}
\acmDOI{10.1145/3831423.3831478}

\copyrightyear{2026}
\acmYear{2026}
\setcopyright{cc}
\setcctype{by}
\acmConference[GI 2026]{Graphics Interface 2026}{June 09--12, 2026}{Waterloo, ON, Canada}
\acmBooktitle{Graphics Interface 2026 (GI 2026), June 09--12, 2026, Waterloo, ON, Canada}
\acmDOI{10.1145/3831423.3831478}
\acmISBN{979-8-4007-2666-8/2026/06}

\usepackage{booktabs} %
\usepackage{graphicx}
\usepackage{subcaption}
\definecolor{light-gray}{HTML}{D3D3D3}
\usepackage{listings}
\usepackage{xcolor}

\usepackage{exii-macros}

\usepackage{booktabs}

\makeatletter
\g@addto@macro\normalsize{%
  \setlength\abovedisplayshortskip{-9pt}
  \setlength\belowdisplayshortskip{3pt}
}
\makeatother

\begin{document}

\tolerance=400 

\title[Sketch Bug]{Sketch Bug: Using Sketch-Based Input for Interactive Code Debugging}

\author{Helen Weixu Chen}
\orcid{0009-0008-1384-6781}
\affiliation{%
  \institution{Cheriton School of Computer Science\\
  University of Waterloo}
  \country{Waterloo, ON, Canada}
}
\email{w352chen@uwaterloo.ca}

\author{Daniel Vogel}
\orcid{0000-0001-7620-0541}
\affiliation{%
  \institution{Cheriton School of Computer Science\\
  University of Waterloo}
  \country{Waterloo, ON, Canada}
}
\email{dvogel@uwaterloo.ca}

\renewcommand{\shortauthors}{Chen and Vogel}

\begin{abstract}
We investigate sketch-like pen input as an alternative way to support execution control in interactive debugging. In our interface, programmers draw lightweight marks to set breakpoints, use symbolic strokes to control execution, and extend strokes into spirals to repeat traversal actions. The prototype combines gesture recognition with Python execution tracing in a conventional editor interface. In a controlled study with 24 programmers, we compared the sketch interface with conventional mouse-and-keyboard input on debugging tasks that required breakpoint placement, step-wise execution, and runtime state inspection. The results show that sketch-like input can support these execution-control tasks, while also introducing challenges in precision, recognition, and gesture recall. Our findings suggest that pen input is most promising where debugger interactions benefit from spatial grounding or continuous movement, rather than as a wholesale replacement for conventional debugging controls.

\end{abstract}

\ccsdesc[500]{Human-centered computing~Gestural input}
\ccsdesc[500]{Human-centered computing~Empirical studies in HCI}
\ccsdesc[500]{Software and its engineering~Software maintenance tools}

\keywords{pen-based interaction, interactive debugging, sketch-based interaction, gesture-based input, execution control}

\begin{teaserfigure}
  \includegraphics[width=\textwidth]{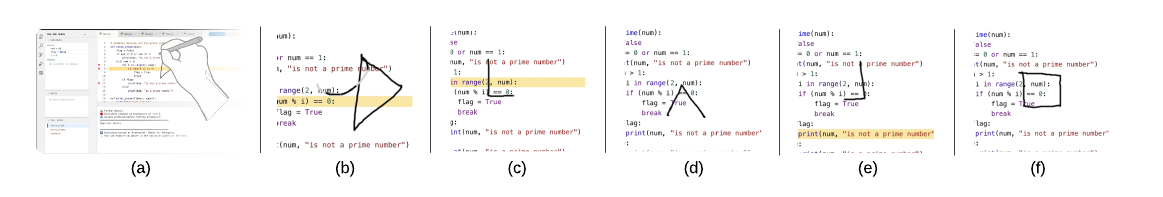}
  \caption{Sketch-like pen gestures: (a) a programmer draws a continue symbol on the canvas; (b) after a 300ms dwell, they add clockwise spirals strokes to repeatedly advance execution; (c) a step into symbol is drawn; (d) a step over symbol is drawn; (e) a step out symbol follows; (f) a stop symbol terminates the debugging session.}
  \label{fig:teaser}
\end{teaserfigure}

\maketitle

\section{Introduction}
Debugging is one of the most cognitively demanding and time-consuming aspects of programming \cite{quinn2022debugging, ko2004six}. Interactive debugging involves many activities, including setting breakpoints, stepping through execution, inspecting runtime state, and relating observed behaviour back to hypotheses about the code \cite{petrillo2016towards}. In practice, a recurring subset of this work centers on lightweight execution-control actions such as placing breakpoints, advancing execution, and restarting a session. In current code editors, these actions are typically carried out through buttons, menus, or keyboard shortcuts.

Prior work has shown that sketching and pen-based interaction can support programming by enabling direct manipulation and integrating input more closely with the artifact being worked on \cite{yen2025code, li2008algosketch, buchanan2014cstutor}. For example, Code Shaping showed how free-form sketches drawn on code can guide AI-mediated edits \cite{yen2025code}, while other systems have used sketching for algorithm design \cite{castelo2022sketching}, execution reasoning \cite{li2008algosketch}, and computing education \cite{buchanan2014cstutor}. In contrast, work on interactive debugging has primarily focused on visualizations, steering mechanisms, and structured controls for understanding program behaviour \cite{yuan2025debug, gray2025interactive, epperson2025interactive}. To our knowledge, sketch-like pen input has not been explored as a way of directly controlling program execution in a debugger.

We argue that pen input is not equally well suited to all aspects of debugging, but it may be promising for a narrower class of debugger interactions. In particular, some debugging actions are spatially anchored to code locations, such as placing or removing breakpoints in the gutter, while others involve repeated execution traversal, such as stepping through many lines of code. These operations may benefit from lightweight marks and continuous strokes drawn directly over the editor, allowing control to remain visually and physically coupled to the code. Rather than proposing pen input as a wholesale replacement for conventional debugging, we explore it as an alternative interaction layer for execution control.

To investigate this opportunity, we introduce \textit{Sketch Bug}, a sketch-like interface for interactive debugging. Programmers can set and clear breakpoints with marks, start or stop a session by drawing symbols, step through execution with simple strokes and shapes, and repeat commands with a spiral gesture. These interactions are rendered as transient ink over a code-editor interface, preserving the familiar structure of a conventional debugger while adding a pen-based control layer. In a within-subject study with 24 programmers, we found that sketch input could compress some repeated control operations into fewer discrete interaction units, especially for breakpoint manipulation and repeated stepping, but also introduced learning, recall, and precision challenges. Our findings suggest that pen input may be most useful not as a replacement for mouse-and-keyboard debugging, but for selected debugging tasks, particularly those involving spatially grounded or repetitive execution-control actions.

This paper contributes:
\begin{itemize}
  \item an exploratory sketch-like pen interface for execution control in interactive debugging; 
  \item empirical findings from a user study examining the strengths and limitations of pen-based debugging interactions;  \item design implications for when pen input appears promising in debugging.
\end{itemize}

\section{Background and Related Work}
We review prior work related to interactive debugging tools as well as sketch- and pen-based interfaces.

\subsection{Interactive Debugging Tools}
Researchers have long explored ways to make debugging more interactive beyond the command-line model of traditional debuggers. Early systems like Whyline demonstrated the potential of query-driven debugging by letting programmers ask "why" and "why not" questions about program execution traces \cite{ko2004designing}. Although powerful, its scope was limited to predefined queries determined by the system. Similarly, VIDA reimagined breakpoint management by recommending candidate breakpoints based on test execution data and visualizing debugging histories \cite{hao2009vida}, lowering the burden of remembering and managing breakpoints. Program slicing has also been widely studied as a technique for isolating code relevant to errors \cite{korel2002pelas}. Tools, such as C-Debug \cite{samadzadeh1993interactive}, extended this idea through dynamic slicing and dicing, allowing programmers to iteratively refine error localization within execution traces through interactive menu-based commands. 

Recent work has further pushed interactivity by connecting debugging to broader contexts. Oddity \cite{woos2018graphical} targeted distributed systems by allowing developers to observe and perturb communication events across executions, although it did not expose a node's internal state. Other visual debugging approaches emphasized synchronized visualizations and interactive parameter adjustment \cite{10.1145/2909480, gray2025interactive}, making complex behaviours more comprehensible through direct manipulation. The debug-gym framework introduced a text-based interactive environment where agents can use tools such as view, rewrite, and pdb to actively investigate and fix bugs, expanding the action and observation space for automated code repair \cite{yuan2025debug}. More recently, AGDebugger supports debugging and steering of multi-agent AI systems through resets, counterfactual exploration, and workflow editing, offering richer control than static prompt engineering or pipeline tools \cite{epperson2025interactive}. 

Across these systems, a common pattern is that interactivity remains tied to predefined queries, recommended breakpoints, or structured menu commands. By contrast, our approach demonstrates how existing interactive debugging commands can be issued through lightweight sketches. 

\subsection{Sketch- and Pen-Based Interfaces}
Researchers have also leveraged sketching in technical domains such as programming and simulation. Yuan et al. developed a pen-based flowchart recognition tool that converted sketched diagrams into C programs helping novices bridge abstract programming constructs with concrete visuals \cite{yuan2008novel}. Similarly, systems like Sim-U-Sketch and AC-SPARC turned hand-drawn diagrams into executable models for circuit analysis or dynamic simulation \cite{stahovich2011pen, gennari2005combining}. These examples demonstrate sketching's potential to lower the threshold for interacting with complex systems, whereas our approach applies the same principle to debugging tasks by mapping sketches directly to runtime operations. 

Another line of work focused on sketching as a medium for prototyping and UI design. Systems like UISKEI allowed users to draw interface elements that were recognized and transformed into fuctional widgets \cite{segura2012combination}, while SketchWizard supported Wizard-of-Oz evaluation of sketch-based interaction \cite{davis2007sketchwizard}. Both projects emphasized recognition pipelines and transient feedback to maintain a natural drawing experience. Our sketch-based approach similarly relies on gesture-like interactions with recognition and feedback, but uses these mechanisms to manage breakpoints and execution flow, situating sketch-based input directly within a live coding environment. 

Sketch input has also been explored as a modality for specifying dynamic behaviours. Motion Doodles allowed animators to draw motion paths for characters \cite{thorne2004motion}, while robot navigation systems used sketched trajectories or landmarks to guide movement \cite{skubic2003sketch}. These applications illustrate how informal strokes can be transformed into precse behaviours. Our approach extends this insight to debugging, treating sketched strokes as direct manipulations of program state rather than annotations or design artifacts. 

Finally, recent work has pushed sketching beyond annotation toward actionable programming commands. Code Shaping allowed programmers to sketch on and around code to guide AI-driven edits, producing high-level modifications such as adding functions or restructuring data flow \cite{yen2025code}. This demonstrates how sketches can tightly couple with code editing but relies on AI assistance to carry out transformations. In contrast, our approach is not AI-assisted: sketches directly map to debugger operations. But we do continue this trajectory of embedding sketch interaction into a code editor, shifting the focus from editing to debugging.

\section{Design Goals}
Our goal was not to replace all aspects of interactive debugging with pen input, but to explore whether sketch-like interaction could support a narrower subset of debugger actions, especially execution control. In designing Sketch Bug, we focused on interactions that are frequent, lightweight, and often tightly coupled to the code surface, such as placing breakpoints, advancing execution, and repeating commands. We were guided by prior work on direct manipulation \cite{yen2025code, Rosenberg2024DrawTalkingBIA} and sketch-based interaction \cite{Kim2024SquidgetsSWA}, as well as concepts from cognitive theories of representation and notation. In particular, we developed the interface around three design goals:

\begin{enumerate} [label=D\arabic*., leftmargin=2.5em, labelsep= 0.8em, itemsep=0.6em]
\item{\textbf{Spatially grounded control.}}  
Debugger actions should remain visually and physically close to the code locations they affect. This goal was informed by the idea of \emph{closeness of mapping} \cite{Zong2020Lyra2DA}: when an operation has an obvious spatial relationship to an object, placing the interaction near that object can reduce semantic distance and make the action easier to interpret.

\item{\textbf{Lightweight visual encoding of command intent.}}  
Pen interactions should use simple marks and shapes that make debugger actions externally visible on the code surface. This goal was informed in part by dual-coding perspectives \cite{Clark1991DualCTA}, which suggest that visual forms can complement abstract symbolic operations. It also reflects concerns from Cognitive Dimensions of Notations about visibility and role-expressiveness \cite{Izquierdo2016CollaboroACA}: the interface should make actions legible without permanently cluttering the workspace.

\item{\textbf{Continuous control for repeated execution traversal.}}  
Repeated debugger actions should be expressible through continuous motion rather than only through many discrete button presses or keystrokes. This goal was informed by direct manipulation principles \cite{Masson2023DirectGPTADA, BeaudouinLafon2000InstrumentalIAA} and by the repetitive nature of many execution-control tasks.
\end{enumerate}

Overall, these goals do not apply equally to every debugger function; rather, they help us examine where sketch-like pen input appears most and least promising within execution control.

\section{Sketch Bug}
We explore how sketch-like pen input can support a narrower subset of interactive debugging actions, especially execution control. Rather than relying on menus, toolbars, or keyboard shortcuts, programmers issue commands by drawing simple strokes directly on a canvas overlaid on the code. These are derived from familiar debugging icons, including triangle for continue, square for stop, caret for step over, and extended with single-stroke variations for more complex operations, such as step into and step out. By turning established debugger semantics into sketchable forms, the goal is to make debugging a more fluid, embodied activity while preserving the rigour of traditional tools (see supplementary video for a demonstration).

\subsection {Interface Design}
The Sketch Bug interface supports the core set of operations of interactive debuggers for session control and execution flow. 

\subsubsection{Session Control}
Session control manages the lifecycle of a debugging session, including starting, stopping, and pausing execution at breakpoints.

\begin{figure}[h]
\centering
\includegraphics[width=\onecolwidth]{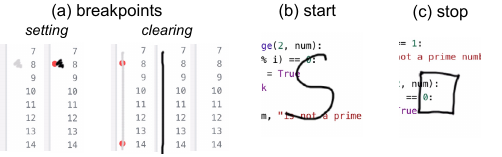}
\caption{Session control interactions: (a) set and remove breakpoints with any mark beside line numbers; (b) start execution with `S' symbol; (c) stop execution with `box' symbol. }
\label{fig:session-gestures} 
\end{figure}

\medbreak
\textit{Breakpoints} – In most code editors, breakpoints appear as red dots in the line number gutter. In our interface, drawing a mark in this area, such as a line, circle, or scribble, is recognized as setting or clearing a breakpoint (\autoref{fig:session-gestures}a). In line with spatially grounded control (D1), breakpoint interactions are drawn directly in the gutter beside the affected line. Drawing a small mark in an empty space in the gutter sets a breakpoint at the corresponding line.  Drawing a mark over an existing breakpoint clears it. This extends to drawing a long vertical stroke across multiple breakpoints to clear them all at once.

\medbreak
\textit{Start/Restart} – Drawing an `S' symbol anywhere in the code editor starts or restarts a debugging session (\autoref{fig:session-gestures}b). To keep command intent lightweight and visible (D2), this operation is expressed as a simple symbolic mark. VS Code separates start (via Run and Debug menu and configuration) and restart, but we unify them into one action. 

\medbreak
\textit{Stop} – The current session is terminated by drawing a rectangle symbol anywhere in the code editor (\autoref{fig:session-gestures}c). Once invoked, execution halts and the session is closed. We chose a rectangle to reflect the conventional stop icon used in the VS Code debugger, keeping the command legible as a compact visible mark on the code surface (D2).

\subsubsection{Execution Flow}

Execution flow controls how program execution advances within and across function boundaries. To support D2, these interactions translate familiar debugger commands into simple symbolic strokes drawn directly on the code surface.

\begin{figure}[h]
\centering
\includegraphics[width=\onecolwidth]{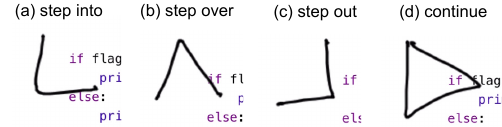}
\caption{Execution flow strokes: (a) step into with `L' shape; (b) step over with caret shape; (c) step out with `mirrored L' shape; (d) continue with `triangle' shape }
\label{fig:execution-gestures} 
\end{figure}

\medbreak
\textit{Step Into} – Drawing an L-shaped stroke (\autoref{fig:execution-gestures}a) advances execution by either stepping to the next line of a function or stepping into the body of a function if the current line is a function call. 
An L-shape was chosen because the downward vertical part is analogous to advancing program execution ``down'' one line in the editor, and the rightward horizontal part connotes diverting ``into'' a function when applicable (since deeper scope in code is represented with increased rightward indentation). 

\medbreak
\textit{Step Over} – Drawing a caret stroke (\autoref{fig:execution-gestures}b) steps to the next line without entering a function call. This means the debugger runs the call to completion or until encountering a breakpoint in the called function. If no breakpoints, execution halts at the next line in the same scope. A caret references `jumping over' with forward progress at the current level.

\medbreak
\textit{Step Out} – Drawing a mirrored L-shape (\autoref{fig:execution-gestures}c) executes the remaining statements in the current function, unwinds the stack, and pauses after the original call. The stroke is the inverse of Step \rev{Into}, with the leftward part suggesting moving out of the current scope.

\medbreak
\textit{Continue} – Drawing a triangular shape resumes execution until the next breakpoint or the program terminates (\autoref{fig:execution-gestures}d). A triangle references the play symbol used in conventional debuggers like VS Code.

\begin{figure}[h]
\centering
\includegraphics[width=\onecolwidth]{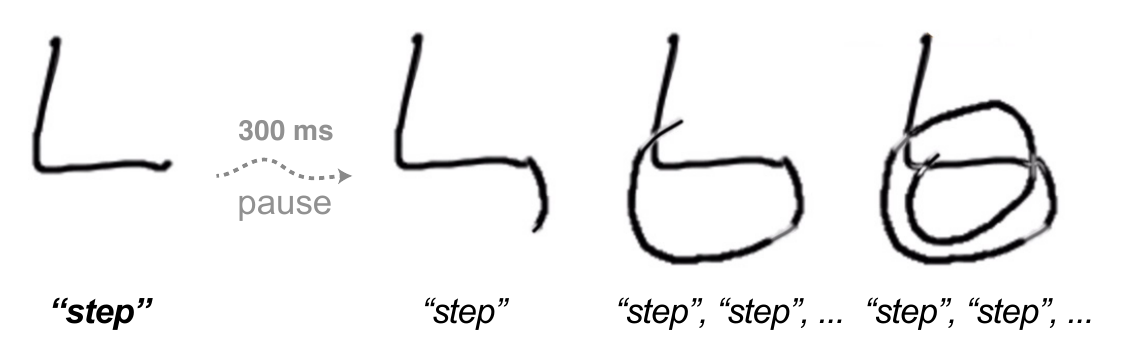}
\caption{Repeating spiral: draw an execution stroke, pause with pen in contact for 300 ms, then draw clockwise spiral to repeat the same command multiple times.  }
\label{fig:repeating-gesture} 
\end{figure}

\medbreak
Guided by D3, all execution flow strokes share a common method to repeat them: 300ms after sketching a base stroke without lifting up, a clockwise spiral repeats the debugging action multiple times (\autoref{fig:repeating-gesture}). 
Rather than requiring a sequence of repeated discrete commands, this design lets users continue traversal through one uninterrupted pen movement. The drawing speed directly controls execution pace: fast movements repeat the action more quickly.
To be specific, the system continuously checks the spiral motion: as long as the pen is moving clockwise, execution advances step by step. The pace scales with the drawing speed but is capped at four steps per second, ensuring that repetition remains predictable. If the pen pauses, no steps are taken.
This is useful when stepping through many lines of code. Action repetition stops when the pen is lifted or the program ends.

The variable panel updates at each step to reflect the changing runtime state, with faster execution producing more rapid updates.

\subsubsection{Feedback}
\rev{Because sketch-based input can be ambiguous, gesture recognition is accompanied by immediate visual and textual feedback. The default pen colour is light gray while drawing. If the input is successfully recognized as a valid sketch interaction, the strokes are converted to black, providing confirmation of recognition. This lightweight feedback reduces uncertainty and helps distinguish between provisional marks and accepted commands.}
Warnings and errors are displayed in the upper-right corner of the interface. For example, attempting to start the debugger without any breakpoints results in a warning message, preventing confusion about why the debugger did not launch.
\rev{Following design recommendations from Code Shaping \cite{yen2025code}, we avoid unnecessary context switching. The system remains in a default pen mode at all times, there is no need to switch to an erasing or colour mode. To maintain a clean interface, all sketches are automatically cleared 500ms after the pen is lifted, so symbols, marks, and strokes do not accumulate and obstruct the code. This design balances the look and feel of sketching with the clarity required for sustained debugging sessions.}

\subsection{Implementation}
\begin{figure*}[tb]
\centering
\includegraphics[width=\twocolwidth]{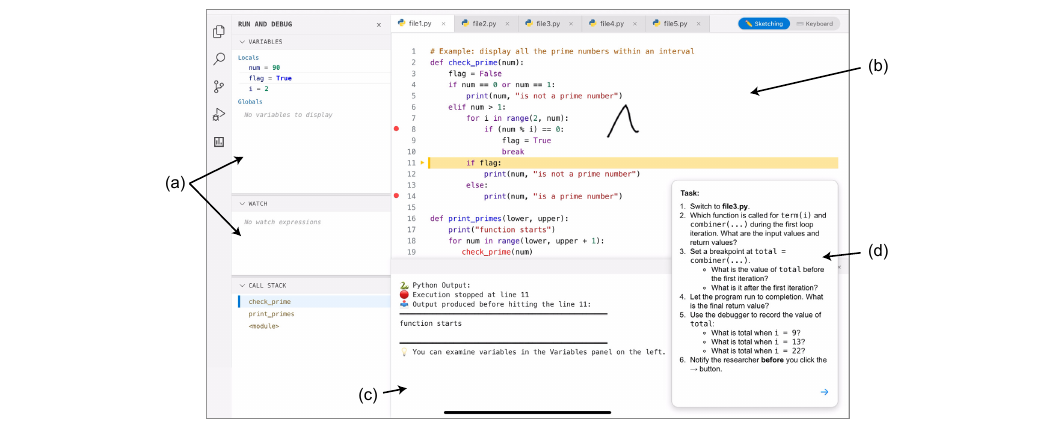}
\caption{Simulated VS Code interactive debugging user interface: (a) debug information panels; (b) code editor with overlaid sketch stroke; (c) debug console output; (d) study instruction panel. }
\label{fig:interface} 
\end{figure*}

Our prototype interface emulates the look and feel of VS Code (\autoref{fig:interface}), a very popular code editor \cite{amann2016study}. The motivation is that developers are familiar with this kind of GUI and can focus on learning the sketch interactions.

The system is implemented as a browser-based application using pure JavaScript, HTML, and CSS. A canvas layer is placed on top of the editor to capture pen or mouse sketches that directly control program execution. As the user sketches, the system records each trajectory as a sequence of points with coordinates and timing information. These sequences are normalized for scale and rotation and recognized using the \$1 Unistroke Recognizer \cite{wobbrock2007gestures}.
Note that for breakpoints and spiral repetition, we use lightweight custom recognizers: any mark in the line number area toggles a breakpoint, while spirals are identified through direct analysis of stroke direction.

A custom execution engine manages program flow line by line. The engine maintains a trace of program states, including the current line number, variable values, and call stack information. Each recognized command invokes a corresponding function that advances or halts this trace to correspond to the desired execution flow (i.e. step into, step over, step out, and continue).
After each advancement, the engine triggers an update cycle that synchronizes the interface: the editor highlights the current line in yellow, the variables panel is redrawn to display updated runtime values, and the call stack view reflects any function entry or return.

All elements of the interface are rendered in the browser and updated dynamically. Because all stroke capture, recognition, execution, and updates run entirely on the client side, the system avoids back-end dependencies and maintains low-latency interaction suitable for both pen-enabled and conventional devices. 
The prototype also has features to run a study, such as the ability to display an instruction panel with task instructions (\autoref{fig:interface}d) and it can be configured to work like a conventional mouse and keyboard interface for comparison.

\section{User Study}
The goal of this study was to examine how sketch-like pen input supported execution-control tasks in interactive debugging, and what tradeoffs emerged relative to conventional mouse-and-keyboard input. 
Using a within-subjects design, participants completed a variations of a debugging task with our sketch interface and with a conventional interface. The study was reviewed and approved by the relevant ethics review board.

\subsection{Participants}
We recruited 24 participants: ages 18 – 44; 14 self-identified as men and 10 as women; 1 participant reported being left-handed. Their programming experience ranged from 1 to 20 years (Mdn = 7, SD = 7.81). All participants indicated prior use of interactive debuggers, such as the VS Code debugger or gdb. On a 5-point scale (1-not at all confident to 5-extremely confident), participants reported relatively high confidence in using debugging tools, including basic and advanced features such as breakpoints, watch expressions, and stepping through call stacks (Mdn = 4, SD = 1.21).
Note that 21 participants took part in person, while 3 participated remotely.

\subsection{Apparatus}
\begin{figure}[h]
\centering
\includegraphics[width=\onecolwidth]{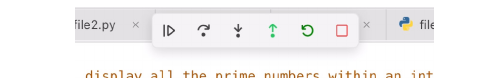}
\caption{Standard debug control buttons for mouse and keyboard condition.}
\label{fig:wimp-toolbar} 
\end{figure}

The Sketch Bug interface can be put into a mouse and keyboard mode. The interface looks and behaves the same except the sketch interactions are disabled and a toolbar with standard debug control buttons is displayed (\autoref{fig:wimp-toolbar}). Standard keyboard shortcuts, like ``F5'', are also enabled.
For in-person sessions, we provided a 14-inch MacBook Pro (Apple M2 Pro) paired with an 11-inch iPad Pro (2nd generation) for sketch input. For remote sessions, participants used their own setup, which required both a personal computer and a tablet (or iPad). The study software logged all interactions with the system, including clicks, sketch interactions, and responses to questionnaires. Each session was audio and screen recorded. 

\subsection{Tasks}
We use a Python debugging task that designed to exercise execution tracing and runtime state inspection, which are common uses of interactive debuggers. 
We avoided syntax errors or trivial mistakes, as modern code editors already flag such issues during editing. Instead, we focused on the types of problems for which interactive debuggers are most often used: tracing program behaviour, monitoring variable changes, and intermediating program state where the output does not meet expectations \cite{ko2004six, latoza2006maintaining, 4016573}. 
These tasks were chosen to elicit the kinds of execution-control actions emphasized by our interface, including placing breakpoints, starting and stopping execution, and stepping through code.
To facilitate a within-subjects design, we created two variations of the debugging task. They have comparable length and difficulty, each involving higher-order functions and iterative loops. The two variations can be found in \autoref{apx: task_variations}.

\textit{Variation 1} was a 16-line program implementing a generic accumulate procedure that combines terms with a specified function. The task required tracing how values were updated across iterations, inspecting the role of the combiner and term functions, and reporting intermediate and final results. This task is adapted from prior work on debugging unexpected code behaviour \cite{suzuki2017tracediff}, as it represents a canonical CS1-style problem that is tractable in length yet enough to elicit meaningful debugging activity.

\textit{Variation 2} was a 14-line program that repeatedly applied an update function until a stopping condition was satisfied. The task required identifying initial variable values, monitoring changes to value across iterations, and examining the role of the update and stop functions. By stepping through execution and observing when termination occurred, participants were able to reason about both intermediate states and the program’s final output. This task emphasized iterative reasoning and conditional control flow, providing a comparable counterpart to the accumulate task.

Each task variation required the participant to complete 8 sub-tasks that involved operations like setting breakpoints, starting and stopping execution, and stepping through code. 
Although the tasks were relatively short, their internal structure was sufficiently complex to ensure that participants could only obtain the correct answers by actively using the debugger.
For example, ``What is the returned value of the function call at the 22nd iteration of the loop?'' 
Sub-tasks were presented sequentially one-by-one on screen. The participant pressed a ``$\rightarrow$'' button when they judged the task variation complete. 

\subsection{Procedure}
After completing a demographics questionnaire, the participant was informed that they would use two different interfaces for interactive debugging tasks. 
The participant began by completing a \textit{tutorial} for the first debugging interface.
This consisted of a sequence of five short video demonstrations (between 26 and 104 seconds), each immediately followed by a practice prompt. The videos presented the features of the interface through abstract examples, and the subsequent practice required participants to try the demonstrated interaction before continuing. This helped participants gradually familiarize themselves with the interface and provided an opportunity to ask questions. The tutorial took approximately 15 minutes. 

Then, the participant completed an assigned variation of the \textit{debugging task} (explained above). When the debugging task was complete, the participant completed a NASA-task load index (NASA-TLX) \cite{hart1988development} to report 5 aspects of perceived workload on a scale from 0 to 20.
Once the debugging task was completed for the first interface, the participant moved to the second interface and completed the same style of tutorial and the other variation of the main debugging task.

Afterwards, there were System Usability Scale (SUS) questionnaires \cite{brooke1996sus} for both interfaces presented together to encourage reflection and comparison. 
Finally, a short open-ended interview was conducted about the participant's experience, including preferences and their overall impression of the design and usability of each interface. The interview questions can be found in \autoref{apx: interview}.
The session lasted approximately one hour, and participants received \$30 in appreciation for their time.

\subsection{Design}

The study used a within-subject design with two interface conditions: \f{sketch} (the Sketch Bug prototype sketching interface) and \f{wimp} (the conventional interface using mouse and keyboard).
The order of interface was counterbalanced along with the assignment of task variation to interface, creating 4 presentation orders that were distributed uniformly across participants.  
This avoided potential confounds of interface order or interactions between interface and task variation.

\subsection{Data Analysis}
We report 95\% confidence intervals (CI) of mean differences \cite{dragicevic2016fair, group2019transparent}. For interpretability, these are reported as $\f{sketch} - \f{WIMP}$. We adopt this convention because sketch-based debugging is the focus of our study, while mouse and keyboard interaction serves as a baseline.
CIs were computed using the studentized bootstrapping method with 10,000 replications, recommended for within-subject study designs with small sample sizes \cite{masson2023statslator, zhu2018assessing}.
To complement these estimates, we also report Wilcoxon signed-rank tests. This non-parametric test is less likely to produce false positives compared to parametric alternatives \cite{bridge1999increasing}. Interview responses were analyzed qualitatively using open coding. We reviewed participants’ comments and iteratively identified recurring ideas related to the perceived strengths, limitations, and appropriate use cases of the interface. Related codes were then grouped into broader categories that informed our qualitative interpretation of the results.

\subsection{Quantitative Results}
\subsubsection{Actions}
\begin{figure}[tb]
	\centering
	\includegraphics[width=\onecolwidth]{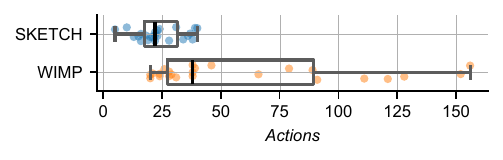}
	\caption{Actions by interface technique.}
	\label{fig:action} 
\end{figure}

An action was defined as the number of strokes in \f{sketch} and the number of mouse clicks or keypresses in \f{wimp}, based on system logs. Because some sketch interactions, such as the repeating spiral, can compress multiple repeated commands into a single stroke, this measure should be interpreted as the number of discrete interaction units rather than as a direct measure of effort. Participants produced fewer such interaction units with \f{sketch} than with \f{wimp} (see \autoref{fig:action}). On average, \f{sketch} required 23.8 actions, whereas \f{wimp} required 60.1, a mean difference of -36.3 (\ci{-59.8}{-19.3}). This difference was significant (\wilcoxonSR{31.5}{.0007}{=}).

\subsubsection{Completion Time, Usability, and Workload}
Participants spent comparable amounts of time completing tasks with the \f{sketch} interface and with mouse and keyboard interface. On average, participants completed tasks in 6.05 minutes with \f{sketch} and 5.20 minutes with \f{wimp}, with no significant difference (95\% CI [-0.34, 2.02], $p=.08$). Although participants received tutorials and practice, some pauses reflected the need to recall gesture-command mappings. This suggests that the observed difference may reflect learning overhead rather than a fundamental limitation of the interaction technique, and performance may improve with greater familiarity over time \cite{li2022memory}. Both interfaces were also perceived as usable. \f{sketch} received a mean SUS score of 72, typically interpreted as ``good,'' while \f{wimp} scored 82, typically considered ``excellent,'' with no significant difference. Similarly, NASA-TLX ratings were comparable across interfaces, with no significant differences observed for mental demand, effort, frustration, physical demand, or perceived performance. Full statistical results are provided in \autoref{apx: stat}.

\subsection{Interview}
\paragraph*{Finding 1: Continuous gestures enable fluid temporal control compared to discrete clicks.}
Fourteen participants (P1, P3, P4, P9, P11, P12, P15, P16, P17, P20, P21, P22, P23, P24) mentioned that the spiral gesture (clockwise stroke) for controlling execution speed provided a superior sense of flow compared to standard interfaces. By mapping the speed of drawing to the processing time of execution, participants felt they had more dynamic control over the debugger. P15 noted, \textit{"I liked the spiral a lot more than having to click continue multiple times... I can speed up or slow down just by changing my hand speed."}. P4 highlighted the physical benefit, stating that drawing spirals was \textit{"less physically tiresome"} than repetitive clicking at a computer. This continuous interaction allowed users to navigate loops and long code blocks more intuitively than static button presses.

\paragraph*{Finding 2: Lack of precision and overshooting are significant usability barriers.}
Twelve participants (P10, P11, P12, P13, P14, P16, P18, P20, P21, P22, P23, P24) reported that while sketching is faster, it lacks the deterministic precision required for granular debugging. The high sensitivity of the spiral gesture often caused users to skip past their intended stopping point, requiring a full session restart. P24 shared their frustration, saying, \textit{"When I'm going very fast... it went 10 steps already. I only need two to three."}. Similarly, P22 noted that \textit{"it's hard to stop in the proper moment sometimes."}. This led many participants to conclude that keyboard input is still superior for tasks requiring line-by-line exactness.

\paragraph*{Finding 3: Recognition failures and attention fragmentation disrupt the debugging flow.}
Eleven participants (P1, P2, P5, P8, P18, P19, P20, P21, P22, P23, P24) felt that gesture recognition issues and system latency created a disjointed experience. Because the system required specific shapes to be drawn in a single stroke, participants often had to divide their attention between the code and the drawing area. P18 explained, \textit{"While you're concentrating here [on the code], if you try and move it [the pen] properly, sometimes you might miss the variable you're actually going for... you have divided attention."}. P2 added that it was \textit{"a bit annoying"} when a gesture did not react as wished, forcing them to re-draw until the stroke turned from grey to black.

\paragraph*{Finding 4: Sketching provides a creative mental break and supports high-level comprehension.}
Eleven participants (P2, P4, P7, P9, P12, P15, P17, P18, P20, P21, P23) described the sketch interface as more \textit{"natural,"} \textit{"creative,"} and \textit{"soothing"} than traditional debugging. Moving from a keyboard to a pen allowed for a psychological shift that supported code understanding rather than just bug fixing. P4 remarked that the action of drawing felt like a \textit{"great mental break"} and a \textit{"creative outlet"} during the stressful process of identifying code errors. P2, who prefers physical writing, mentioned, \textit{"The fact that you bring something from a totally digital environment into physical... it was really soothing for me."}.

\paragraph*{Finding 5: Strong potential for Education, tablet-based debugging, and mobile review.}
There was a strong consensus among participants (P1, P4, P8, P11, P13, P15, P17, P18, P19, P21, P23, P24) that this approach is best suited for education and mobile scenarios. P23, who has extensive experience tutoring coding, stated, \textit{"The sketch mode will be more entertaining and interactive [for kids]... debugging is not like some super professional people do."}. Others highlighted the utility of \textit{"couch debugging"} on an iPad, where a mouse is unavailable. P17 noted that sketching would be \textit{"obviously useful"} for dedicated debugging sessions when \textit{"I don't want to be at my computer."}.

\section{Discussion}
Our results suggest that sketch-like pen input changes debugging not only by introducing a new command modality, but by shifting how execution control is expressed and experienced. The benefits appeared in interactions that were either continuous or spatially grounded, while the costs arose from limited precision, recognition overhead, and the need to recall symbolic mappings.

\textit{Continuous pen input improved the flow of execution traversal, but made precise stopping more difficult.}
Participants often described the spiral gesture as smoother and less tedious than repeated clicking, especially when moving through loops or longer stretches of code. In practice, this interaction seemed most useful when programmers already had a clear traversal goal, for instance, when they wanted to move quickly toward a relevant part of execution and then slow down for closer inspection. At the same time, the same mechanism reduced precision: overshooting and difficulty stopping at the intended moment were among the most common usability barriers. These results point to a design tension for pen-based debugging: continuous input can support fluid, goal-directed traversal, but it remains fragile for exact line-by-line control.

\textit{Spatially grounded interactions appeared more compelling than symbolic command strokes.}
Not all gestures contributed equally. Breakpoint manipulation and repeated traversal benefited from being tied either to the code surface or to continuous pen motion. However, several symbolic command gestures functioned more like pen-based shortcuts: they made command intent visible, but still required memorization and accurate recognition. This distinction helps clarify what sketch-like interaction contributed in our prototype. Its strongest value did not come from replacing every debugger command with a drawn symbol, but from supporting actions that had a spatial or temporal structure.

\textit{Sketching changed the felt experience of debugging, but recognition overhead fragmented attention.}
For some participants, pen input made debugging feel more physical, creative, and less monotonous. Several described it as a mental break from conventional clicking, suggesting that sketching altered the rhythm of debugging as an activity, not only the mechanics of issuing commands. However, this experiential benefit came with a cost. Because users had to monitor both the code and the recognizability of their strokes, attention was often divided between reasoning about execution and managing input production. Recognition failures and redraws further disrupted this flow. This suggests that visible, transient command marks can make debugger actions feel more immediate, but only when the recognition layer remains lightweight and dependable.

\textit{The strongest fit may lie in selected contexts rather than full replacement.}
Participants most often imagined this interaction style being useful in situations that emphasized engagement, portability, or dedicated execution tracing, such as instructional settings, tablet-based debugging, or short debugging sessions away from a traditional desk setup. We view these not as validated deployment claims, but as clues about where pen input may add the most value. Rather than serving as a wholesale substitute for mouse-and-keyboard debugging, sketch-like interaction may be most promising in contexts where its spatial, visible, and continuous qualities matter more than maximum precision.

\subsection{Limitations and Future Work}
Our tasks were designed to foreground breakpoint placement, step-by-step execution, and runtime state inspection. Although they required execution reasoning, they did not capture other aspects of real-world debugging, such as navigating across multiple files \cite{Khan2025KodeziCAA}, searching within larger codebases \cite{Li2026AGTA}, or comparing more complex program states \cite{Khan2025KodeziCAA}. Accordingly, our findings speak most directly to execution-control interactions rather than to the full range of debugging practice. Future work should examine how sketch-like interaction performs in broader debugging settings, including tasks that require navigation across larger projects and more complex state inspection.

Continuous pen-based control also introduced limitations in precision and recoverability. As our interview findings showed, overshooting, recognition sensitivity, and the lack of undo all weakened participants’ sense of control, especially during repeated traversal. These issues suggest several concrete directions for future work, including more stable stopping thresholds, adjustable repetition speeds, previews of repeated execution, and lightweight undo or rollback support.

Finally, our prototype focused on core execution-control functions. While this scope was appropriate for an initial investigation, debugging also involves forming hypotheses, tracking expectations, and externalizing interpretations of program behaviour \cite{ko2004designing, Vrseda2012ASFA}. A next step is to explore whether pen input can support these reasoning activities more directly, for example, by allowing programmers to annotate expected values, mark suspected execution paths, or connect pen-based notes to runtime state and execution traces.

\section{Conclusion}
We presented Sketch Bug, a sketch-based interface for execution control in interactive debugging. In a controlled user study, we found that sketch-like pen input could support breakpoint manipulation and execution traversal, while also introducing challenges in precision, recognition, and gesture recall. Our findings suggest that pen input is promising where debugger interactions benefit from spatial grounding or continuous movement, rather than as a wholesale replacement for conventional mouse-and-keyboard control. More broadly, this work highlights opportunities to rethink how debugging commands are expressed and experienced on the code surface.

\begin{acks}
We thank Dr. Amy J. Ko for her early guidance on the Design Goals section and the anonymous reviewers for their feedback, which helped improve our work. We also thank our participants for their time and insights. This work was made possible by 
NSERC Discovery Grant 2024-03827 and  
Canada Foundation for Innovation Infrastructure Fund 33151 ``Facility for Fully Interactive Physio-digital Spaces''.  
\end{acks}

\bibliographystyle{ACM-Reference-Format}
\bibliography{_references.bib}

\appendix
\makeatother
\clearpage
\renewcommand\thefigure{\thesection.\arabic{figure}}
\setcounter{figure}{0}
\appendix
\section{Task Variations}
\label{apx: task_variations}
\subsection{Variation 1}
\begin{lstlisting}
def accumulate(combiner, base, n, term):
    total = base
    i = 1
    while i <= n:
        total = combiner(i, term(i))
        i = i + 1
    return total

def add(a, b):
    return a + b

def identity(x):
    return x

accumulate(add, 0, 25, identity)
\end{lstlisting}

\begin{enumerate}
    \item During the first loop iteration, which functions are called for \texttt{term(i)} and \texttt{combiner(...)}? What are their input values and return values?

    \item Set a breakpoint at \texttt{total = combiner(...)}.

    \item What is the value of \texttt{total} before the first iteration?

    \item What is the value of \texttt{total} after the first iteration?

    \item Let the program run to completion. What is the final return value?

    \item Use the debugger to record the value of \texttt{total}:
    \begin{itemize}
        \item What is \texttt{total} when \texttt{i = 9}?
        \item What is \texttt{total} when \texttt{i = 13}?
        \item What is \texttt{total} when \texttt{i = 22}?
    \end{itemize}
\end{enumerate}

\subsection{Variation 2}
\begin{lstlisting}
def apply_until(stop_fn, update_fn, initial):
    value = initial
    while not stop_fn(value):
        value = update_fn(value)
    return value

def greater_than_100(x):
    return x > 100

def double_plus_one(x):
    return 2 * x + 1

apply_until(greater_than_100, double_plus_one, 1) 
\end{lstlisting}

\begin{enumerate}
    \item Set a breakpoint at the first line inside \texttt{apply\_until()}:
    \texttt{value = initial}

    \item Run the program until it hits the breakpoint. Then answer:
    \begin{itemize}
        \item What is the value of \texttt{initial}?
        \item What functions were passed as \texttt{stop\_fn} and \texttt{update\_fn}?
        \item What is the initial value of \texttt{value}?
    \end{itemize}

    \item Restart the debugger. Step Over until you hit the loop guard, i.e., \texttt{while not stop\_fn(value):}, for the second time.
    \begin{itemize}
        \item What is the new value of \texttt{value}?
    \end{itemize}

    \item When \texttt{value = 63}, step into the function call.
    \begin{itemize}
        \item What is the function name?
        \item What is the input?
        \item What is the return value?
    \end{itemize}

    \item What is the return value of the \texttt{apply\_until} call?
\end{enumerate}

\section{Interview Questions}
\label{apx: interview}
\begin{enumerate}
    \item How did using sketching compare to how you typically interact with a debugger?
	\item Were there moments when using sketches felt especially helpful or intuitive?
	\item Were there moments when using sketches felt especially challenging?
    \item How did using a pen or drawing gestures affect your experience? 
	\item If you could change or add new functionalities for sketches, what would you most like to have? 
	\item In what scenarios do you think this sketch-based debugging approach has the most potential for widespread use?
	\item Is there anything you’d like to share?
\end{enumerate}

\section{Statistical Results}
\label{apx: stat}
\begin{table}[h]
\centering
\caption{Workload comparisons between \f{Sketch} and \f{WiMP}. Mean differences are reported as \f{WiMP} $-$ \f{Sketch}.}
\begin{tabular}{lccc}
\hline
\textbf{Measure} & \textbf{95\% CI} & \textbf{Wilcoxon ($W$, $p$)} \\
\hline
Mental Demand    & [-2.898, 0.880]  & $W = 102.500,\ p = 0.4341$ \\
Physical Demand  & [-2.657, 3.206]  & $W = 113.500,\ p = 0.9445$ \\
Effort           & [-3.218, 2.078]  & $W = 116.000,\ p = 0.7325$ \\
Performance      & [-0.002, 2.509]  & $W = 53.000,\ p = 0.0888$ \\
Frustration      & [-2.622, 2.301]  & $W = 104.000,\ p = 0.9701$ \\
\hline
\end{tabular}
\end{table}

\end{document}